\documentclass[a4paper,fleqn,usenatbib,useAMS]{mnras}
\usepackage[T1]{fontenc}
\usepackage{ae,aecompl}
\usepackage{ogonek}
\usepackage{amstext}
\usepackage{url}
\usepackage{graphicx}	
\usepackage{amsmath}	
\usepackage{amssymb}	
\usepackage{color}

\title[Escape and trapping of rays from COs]{Escape and trapping of 
low-frequency gravitationally lensed rays by compact 
objects within plasma}

\author[Adam Rogers]{Adam Rogers\thanks{E-mail: rogers@physics.umanitoba.ca}
\\
Department of Physics and Astronomy, University of Manitoba, Winnipeg, MB R3T 2N2, 
Canada\\
}

\date{Accepted 2016 November 1. Received 2016 October 31; in original form 2016 August 23.}

\volume{465}
\pagerange{2151--2159} \pubyear{2017}

\begin{document}
\label{firstpage}
\pagerange{\pageref{firstpage}--\pageref{lastpage}}
\maketitle

\begin{abstract}
We consider the gravitational lensing of rays emitted by a compact object (CO) 
within a distribution of plasma with power-law density $\propto r^{-h}$. For 
the simplest case of a cloud of spherically symmetric cold non-magnetized 
plasma, the diverging effect of the plasma and the converging effect of 
gravitational lensing compete with one another. When $h<2$, the plasma effect 
dominates over the vacuum Schwarzschild curvature, potentially shifting the 
radius of the unstable circular photon orbit outside the surface of the CO. 
When this occurs, we define two relatively narrow radio-frequency 
bands in which plasma effects are particularly 
significant. Rays in the escape window have $\omega_{0} < \omega \leq 
\omega_{+}$ and are free to propagate to infinity from the CO surface. To a 
distant observer, the visible portion of the CO surface appears to shrink as the 
observed frequency is reduced, and vanishes entirely at $\omega_{0}$, in excess 
of the plasma frequency at the CO surface. We define 
the anomalous propagation window for frequencies $\omega_{-}< \omega \leq 
\omega_{0}$. Rays emitted from the CO surface within this frequency range are 
dominated by optical effects from the plasma and curve back to the surface of 
the CO, effectively cloaking the star from distant observers. We conclude with 
a study of neutron star (NS) compactness ratios for a variety of nuclear matter 
equations of state (EoS). For $h=1$, NSs generated from stiff EoS 
should display significant frequency dependence in the EW, and lower values of 
$h$ with softer EoS can also show these effects.
\end{abstract}

\begin{keywords}
gravitation -- plasmas -- stars: neutron -- pulsars: general
\end{keywords}

\section{Introduction}
\label{sec:intro}

Gravitational lensing and plasma effects were first considered in the 
literature by \citet{synge}, and later applied by \citet{solarPlasmaDerivation} 
and \citet{solarPlasma}, who calculated the effect of the solar plasma on the 
deviation of background starlight. The combination of gravitational lensing and 
plasma effects were significantly expanded on by \citet{perlickGR} and 
\citet{BKT09}, leading to studies that range from the scale of stellar mass 
black holes in X-ray binaries \citep{BKT10} to galaxy-scale lenses 
\citep{mao14}. This work has been extended to include rotation using the Kerr 
metric \citep{plasmaLensingKerr}. The effect of plasma on the shadows of black 
holes in general relativity has also been explored \citep{shadow2, shadow1, shadow4}, as well 
as near exotic objects \citep{shadow3, shadow5} and in $f(R)$ gravity 
\citep{shadow6}. For a detailed overview of gravitational lensing effects 
within distributions of plasma, we refer to the thorough review provided by 
\citet{review}.

The plasma frequency $\omega_\text{e}$ defines a low-frequency cutoff for 
electromagnetic wave propagation. The index of refraction $n(\omega,r)$ for 
radiation below this frequency is imaginary, which indicates that such 
low-frequency radiation is absorbed by the plasma. Generally the effect of 
plasma on ray trajectories near the compact object (CO) is expressed as an 
additional term in the effective potential. \citet{kulsrudLoeb} showed that a 
constant plasma density provides an effective mass for light rays through this 
potential term. However, inhomogeneous distributions of plasma can generate 
more interesting effects \citep{TBK13}. \citet{rogers2015}
considered rays passing a massive object sheathed in a distribution of cold, 
non-magnetized plasma with a power-law density distribution $r^{-h}$ for 
integer $0<h<3$, and presented detailed calculations of the effect of the 
plasma with $h=3$ on the pulse profiles of a central neutron star (NS) with two 
emitting polar caps. Due to the dipolar magnetic field of the NS, the $h=3$ 
case was primarily investigated. 

In this work, we investigate the consequences of low-frequency 
emission from a CO sheathed in plasma with $h<2$. For this case, the plasma term 
in the 
effective potential drops off more slowly than the Schwarzschild vacuum term. 
When this occurs the diverging-lens behaviour of the plasma dominates for 
low-frequency rays, such that a potential maximum exists external to the stellar 
radius $R$ for sufficiently compact stars. Thus, low-frequency radiation is 
affected by turning points of the effective potential, causing rays which would 
usually escape the surface of the CO in the vacuum case 
to turn back towards the stellar surface. A wide range of other novel behaviour 
exists for rays that escape the CO surface with frequencies $\omega_{0} < 
\omega 
\leq \omega_{+}$. We define this frequency range as the escape window (EW), and 
the trapped rays with $\omega_{-} <\omega \leq 
\omega_{0}$ are within the anomalous propagation window (APW). Ray trapping in 
the APW is analogous to the anomalous propagation of radio signals in the 
Earth's atmosphere, and has profound effects on the appearance 
of a CO to distant observers, such as reducing the 
apparent radius of a CO as a function of frequency. In fact, the CO appears to 
vanish entirely
as the observed frequency approaches $\omega_0$, effectively `cloaking' the 
stellar surface from a distant observer when 
observed at low frequencies. These frequency windows do not exist when the 
plasma density drops off more rapidly than the 
vacuum term in the effective gravitational potential for null geodesics in the 
Schwarzschild space--time. 

We review the details of the geometrical optics 
approach to gravitational lensing within a plasma distribution in Section 
\ref{sec:theory}, provide the details of the frequency windows in Section 
\ref{sub:Freq} and derive the maximum impact parameter as a function of 
frequency in Section \ref{sub:impact}. We discuss our results in Section 
\ref{sec:discussion} and summarize our conclusions in Section 
\ref{sec:conclusions}.

\section{Theory}
\label{sec:theory}

Previous work \citep{rogers2015} examined modifications to ray trajectories 
introduced by a 
simple choice of dispersion relation in the case of spherical symmetry, 
following 
the general approach and notation of \citet{TBK13}. We will briefly review the 
assumptions behind this physical scenario here. 

We make use of the Schwarzschild metric to describe the space--time surrounding 
a 
spherically symmetric CO,
\begin{equation}
\text{d}s^2=-A(r)\text{d}t^2+\frac{\text{d}r^2}{A(r)}+r^2\left( \text{d}\theta^2 + 
\sin^2\theta \text{d}\phi^2 \right)
\end{equation}
with the metric function
\begin{equation}
A(r)=1-\frac{r_\text{g}}{r}
\end{equation}
and abbreviating $r_\text{g}=2M$. In these definitions and for the rest of this 
paper, we use 
units which have 
$G=c=\hbar=1$. Let us assume the index of refraction for a cold plasma
\begin{equation}
n^2(r, \omega)=1-\frac{\omega_e^2}{\omega^2},
\label{nDef}
\end{equation}
with plasma frequency
\begin{equation}
\omega_e^2(r)=\frac{4 \upi e^2 N(r)}{m}
\end{equation}
that has a plasma particle charge and mass given by $e$ and $m$, respectively, 
as well 
as the plasma number density $N(r)$. An often used
plasma density in the literature is a radial power-law \citep{TBK13}, such that
\begin{equation}
N(r)=\frac{N_0}{r^h}.
\end{equation}
This puts the plasma frequency into a simple form,
\begin{equation}
\omega_e^2(r)=\frac{k}{r^h},
\end{equation}
where $k$ is a constant and $h$ is the power-law exponent. For simplicity, we 
set the constant $k=1$ in our numerical calculations. Throughout the text we 
refer to frequencies detected by a distant observer with the subscript 
$\infty$, 
from the redshift relation
\begin{equation}
\omega(r)=\frac{\omega_\infty}{A(r)^{1/2}}
\label{effRedShift}
\end{equation}
and do not explicitly state the radial dependence of $\omega$. The Hamiltonian 
for rays in curved space--time and under the effects of an optical medium with 
index of refraction \citep{synge} using cold non-magnetized plasma is
\begin{equation}
H(x^i, p_i)=\frac{1}{2}\left( g^{ij}p_i p_j + \omega_e^2 \right) = 0.
\end{equation}
The equations of motion are given in terms of an arbitrary curve parameter 
$\lambda$:
\begin{equation}
\frac{\text{d}x^i}{\text{d} \lambda} = \frac{\upartial H}{\upartial p_i} = g^{ij} 
p_j
\label{hamiltonianX}
\end{equation}
\begin{equation}
\frac{\text{d}p_i}{\text{d} \lambda} = - \frac{\upartial H}{\upartial x^i} = 
-\frac{1}{2}g^{jk}_{,i} p_j p_k - \frac{1}{2} \left( \omega_\text{e}^2 
\right)_{,i}.
\label{hamiltonianP}
\end{equation}
From these relationships equation \ref{hamiltonianP} immediately gives
\begin{equation}
\frac{\text{d} p_t}{\text{d}\lambda} = \frac{\text{d} p_\phi}{\text{d}\lambda} 
= \frac{\text{d} p_\theta}{\text{d}\lambda}=0.
\label{derivsP}
\end{equation}
The vanishing derivatives show that the $t$, $\theta$ and $\phi$ components 
must be constant. By spherical symmetry, we choose to work in the equatorial 
plane so take $\theta=\upi/2$. The $t$ and $\phi$ quantities are interpreted as 
the energy and angular momentum of a ray,
\begin{equation}
p_t = - E =- \omega_\infty
\label{pT}
\end{equation}
and
\begin{equation}
p_\phi = L = \omega_\infty b,
\label{pPhi}
\end{equation}
where $b$ is the impact parameter of an escaping ray. Finally, the derivative 
of the radial momentum is
\begin{equation}
\frac{\text{d} p_r}{\text{d}\lambda}= - \frac{M}{r^2 A^2(r)} E^2 - 
\frac{M}{r^2} p_r^2 + \frac{L^2}{r^3} + \frac{1}{2} \frac{kh}{r^{h+1}}.
\end{equation}
We can find an expression for the radial momentum $p_r$ directly from the 
vanishing of the Hamiltonian, such that
\begin{equation}
p_r= \pm \frac{L}{A(r)} \left[ \frac{E^2}{L^2} - A(r)\left( \frac{1}{r^2} + 
\frac{1}{L^2} \frac{k}{r^h} \right) \right]^{1/2}
\label{pR}
\end{equation}
and the sign of $p_r$ depends on the ray trajectory, with positive indicating 
an 
outgoing ray and negative for an incoming ray. The coordinate derivatives using 
equation 
\ref{hamiltonianX} give
\begin{equation}
\frac{\text{d} t}{\text{d}\lambda} = \frac{E}{A(r)}
\end{equation}
\begin{equation}
\frac{\text{d} r}{\text{d}\lambda} = A(r) p_r
\label{dr}
\end{equation}
\begin{equation}
\frac{\text{d} \phi}{\text{d}\lambda} = \frac{L}{r^2}
\end{equation}
and working in the equatorial plane implies
\begin{equation}
\frac{\text{d} \theta}{\text{d}\lambda}=0.
\label{dThetadLambda}
\end{equation}
For an initial position $x_\text{initial}=(t, r, \theta, \phi)$, ray energy $E$ 
and angular momentum $L$, as well as a choice of sign for $p_r$, equations 
\ref{derivsP} to \ref{dThetadLambda} form a system of equations that are easily 
solved 
using an integration scheme for second-order ordinary differential equations, such as the fourth-order 
Runge--Kutta method. The solution to 
this system gives points along the path of a ray trajectory launched from 
$x_\text{initial}$ in the direction of our choice. We stop integration when the 
ray returns to the surface of the CO ($r < R$), or when it escapes to a 
sufficiently great distance ($r \geq 100 R$).

The trajectory of a ray near the CO is described in terms of an effective 
potential. We denote $\dot{r}=\text{d}r/\text{d} \lambda$, and use equations 
\ref{pR} and \ref{dr} to give an energy conservation equation
\begin{equation}
\dot{r}^2= E^2-V_\text{eff}(r)
\label{Econs}
\end{equation}
with
\begin{equation}
V_\text{eff}=\left(1-\frac{2M}{r}\right)\left[\frac{L^2}{r^2}+\frac{k}{r^h}
\right],
\label{VDef}
\end{equation}
where the vacuum contribution is the first term in the square brackets and the 
second term is entirely due to the plasma. Setting the square of the ray energy 
equal to the effective potential gives the propagation condition
\begin{equation}
\omega_\infty\geq\sqrt{V_\text{eff}(r)}
\label{propCond}
\end{equation}
which must be satisfied at every point along the trajectory for rays to escape 
from the surface of the CO and 
travel to infinity. The impact parameter of an escaping ray with asymptotic 
frequency $\omega_\infty$ is 
\begin{equation}
b(r)=\frac{rn(r)}{A(r)^{1/2}}\sin(\delta).
\label{bDef}
\end{equation}
This ray will intersect the radial normal vector $\hat{\mathbfit{r}}$ at $r$ with 
an angle $\delta$. A ray passing $r$ with a grazing incidence $\delta=\upi/2$ 
and escaping to reach an observer at infinity defines the maximum impact 
parameter, 
$b_\text{max}$ at a particular frequency. All rays that pass $r$ with impact 
parameter $b<b_\text{max}$ will intersect $\hat{\mathbfit{r}}$ at smaller angles.

In analogy to equation \ref{effRedShift}, we use the effective redshift formula 
to define the plasma frequency at infinity,
\begin{equation}
\omega_{\infty 
\text{e}}=A(r_\text{max})^{1/2}\omega_{\text{e}}(r_\text{max}),
\label{asymPlasmaFreq}
\end{equation}
where $r_\text{max}$ is the position of the maximum density along the path of a 
given ray. The rays that propagate through plasma that are visible to distant 
observers require $\omega_{\infty}>\omega_{\infty \text{e}}$ everywhere along 
their 
trajectory. Note that the plasma frequency at infinity is equivalent to the 
square root of the effective potential (equation \ref{VDef}) with $L=0$. This 
means that radially directed rays with sufficiently low asymptotic frequency 
will be absorbed since the index of refraction $n(\omega,r)$ vanishes for rays 
at the asymptotic plasma frequency. These rays have absorption points 
rather than turning points. However, rays with finite $L$ at these same 
frequencies can propagate freely since they remain above the plasma frequency 
cutoff at all times due to the contribution of $L$ to the effective potential.

\subsection{Frequency windows}
\label{sub:Freq}

For plasma density distributions with power-law index $h<2$, the propagation of 
low-frequency rays is complicated by the presence of a turning point that can 
be external to the stellar surface. The dependence between the radius of the 
photon sphere and $h$ was discussed by \citet{shadow1} in their work on black 
hole shadows in plasma. However, the effect has additional implications for 
objects with surfaces that produce low-frequency emission.

We define the EW as a range of frequencies $\omega_{0} < \omega 
\leq \omega_+$ within which rays are strongly affected by the plasma, but are 
still free to escape to an observer at infinity. At $\omega_+$, the effective 
potential allows a circular orbit at the exact surface of the CO for the 
maximum 
impact parameter given by equation \ref{bDef} with $r=R$. This condition 
remains 
valid for higher frequencies $\omega_+ < \omega$ as these have circular orbit 
radii within the stellar surface. At the lower limit of the EW only the 
fiducial ray with frequency $\omega_0$ escapes to a distant observer.

To describe these effects quantitatively, let the circular orbit radius be 
called $r_\text{c}$, the position of the potential maximum where the derivative 
of the effective potential vanishes. This condition gives the critical angular 
momentum required for a circular orbit in terms of $h$,
\begin{equation}
L_\text{c}^2(r_\text{c}) = \frac{ Mr_\text{c}^2 }{( r_\text{c}-3M)}\frac{ k }{ 
r_\text{c}^{h-1} }\left( \frac{h+1}{r_\text{c}} - \frac{h}{2M} \right),
\label{LcDef}
\end{equation}
with $L_\text{c}=\omega_\infty b_\text{c}$ in terms of the observed frequency 
at infinity, and we write the critical impact parameter as 
$b_\text{c}=b(r_\text{c})$. The upper frequency limit of the EW is found when 
the circular orbit and stellar surface coincide at $r_\text{c}=R$, with the 
corresponding angular momentum $L_+=L_\text{c}(r_\text{c})=L_\text{c}(R)$, and 
$b_\text{c}=b(r_\text{c})=b(R)$. Above this limit, $L>L_+$, the circular orbit 
is within the stellar surface and all rays are generally free to escape to 
infinity with maximum impact parameter $b(R)$. For this reason, we will restrict 
ourselves to cases for which $L \leq 
L_+$, in which ray paths are significantly altered from their vacuum behaviour. 
The angular momentum $L_+$ can be used to define a corresponding asymptotic 
frequency which we call the upper limit of the EW, $\omega_{\infty+}$. With the 
effective 
potential (equation \ref{VDef}), and using the equality in the propagation 
condition (equation 
\ref{propCond}), we find an expression for the corresponding asymptotic 
frequency
\begin{equation}
\omega_{\infty+}=\left(1-\frac{2M}{R}\right)\left[\left(1-\frac{h}{2}
\right)\frac{k}{R^{h-1}}\frac{1}{(R-3M)}\right]^{1/2}.
\label{w+}
\end{equation}
From this expression, we see that the upper frequency limit vanishes when $h=2$, 
and real values of the upper frequency limit exist only for plasma 
distributions with $h<2$.

Rays with frequencies $\omega<\omega_+$ can enter a circular orbit outside the 
CO surface. The presence of the unstable circular orbit external to 
the stellar surface implies a maximum in the potential. Therefore a ray 
escaping the CO surface must have energy in excess of this maximum to escape 
and 
avoid a turning (absorption) point in the trajectory. Thus, the effective 
potential maximum 
defines the lower limit of the EW, $\omega_{0}$, which a distant observer 
detects as an asymptotic frequency
\begin{equation}
\omega_{\infty\text{0}}=\sqrt{V_\text{eff}(r_\text{c})}.
\label{w0}
\end{equation}
This is the threshold frequency for rays to escape the stellar surface.

The existence of the EW provides a constraint on the maximum compactness ratio 
of a CO, $R/r_\text{g}$. Let us consider a radially directed ray 
($\delta=0$) that is at the threshold of escape. Equation \ref{LcDef} vanishes 
for a ray that is radially directed. With this condition the position of the 
potential maximum for general $h$ is
\begin{equation}
r_\text{c} = r_\text{g} \left( \frac{h+1}{h} \right).
\label{critRad}
\end{equation}
The existence of a turning point at or external to the CO surface $r_\text{c} 
\geq R$ is necessary for the existence of the EW. Let us set $r_\text{c}=R$ for 
the most conservative limiting case. Combined with the $h<2$ condition implied 
by equation \ref{w+}, we find a minimum compactness ratio
\begin{equation}
\frac{R}{r_\text{g}} \geq \frac{3}{2}.
\label{compressionRatio}
\end{equation}
This compactness ratio is a realistic lower bound for NSs, which we will 
discuss further in Section \ref{sec:discussion}. 

Finally, we also note the simplification introduced for radially directed rays 
at the minimum EW frequency $\omega_{0}$. Using equation \ref{critRad} in 
equation \ref{w0} gives the lower EW limit for a radially directed ray,
\begin{equation}
\omega_{\infty0}=\left[\frac{k}{h+1}\left(\frac{h}{2M(h+1)}\right)^h\right]
^{1/2}.
\label{w0h}
\end{equation}
In the specific case $h=1$ and $r_\text{c}=4M$,
\begin{equation}
\omega_{\infty0}=\sqrt{\frac{k}{8M}}.
\end{equation}
This is the minimum frequency limit for rays to escape along a radial 
trajectory, and equals the asymptotic plasma frequency (equation 
\ref{asymPlasmaFreq}).

For rays with frequencies lower than the EW, we define the APW, in which the diverging effect of the plasma dominates ray 
propagation. The APW is the range of frequencies given by $\omega_{-}< \omega 
\leq \omega_{0}$. Since $r_\text{c}>R$ implies a potential maximum of 
$V_\text{eff}(r_\text{c})=\omega_{\infty0}^2$, the APW is bounded from below by
\begin{equation}
\omega_{\infty-}=\sqrt{V_\text{eff}(R)}.
\end{equation}
Rays in the APW that are emitted by the CO have frequencies larger than the 
local plasma frequency for radii $R<r$, but below the asymptotic plasma 
frequency 
at $r_\text{c}$ required for escape. Thus, these rays are effectively trapped 
in the environment of the CO. Rays in the APW reach a maximum distance 
$r_{-}<r_\text{c}$ where they encounter a turning point due to the potential 
barrier and subsequently reverse direction to curve back to the CO surface. 
Rays external to the circular orbit radius with frequencies in the APW that 
approach the CO from infinity reach a minimum turning distance $r_{+}$ and are 
reflected away from the CO by the potential boundary, returning to infinity. To 
find the external turning radius as a function of impact parameter, let us 
consider the polynomial expression for the turning points \citep{rogers2015},
\begin{equation}
r^{h+1}-b^2r^{h-1}+2Mb^2r^{h-2}-\frac{k}{\omega_\infty^2}r+2M\frac{k}{
\omega_\infty^2}=0.
\label{turningPtDef}
\end{equation}
For a given $b$, the solutions of equation \ref{turningPtDef} give the external turning points $r_{+}$. For rays emitted from the surface of the CO in 
the APW, we substitute 
$b=L/\omega_\infty$ to find the corresponding interior turning points $r_{-}$ 
in terms of $L$. The simplest example is a radially directed ray. 
Using $h=1$ and setting $L=0$ eliminates the $b$ terms, giving
\begin{equation}
r_{\pm}=\frac{k}{2\omega_\infty^2}\left(1\pm\sqrt{1-\frac{8M\omega_\infty^2}{k}}
\right).
\label{rpm}
\end{equation}
However, as seen from the effective potential and equation 
\ref{asymPlasmaFreq}, radially outgoing rays experience absorption at the 
turning points due to the plasma frequency. Thus, rays are absorbed at $r_\pm$, 
but all other rays with non-vanishing $L$ are free to propagate.

\begin{figure}
\includegraphics[bb=  187 200 392 598, clip, scale=0.88]{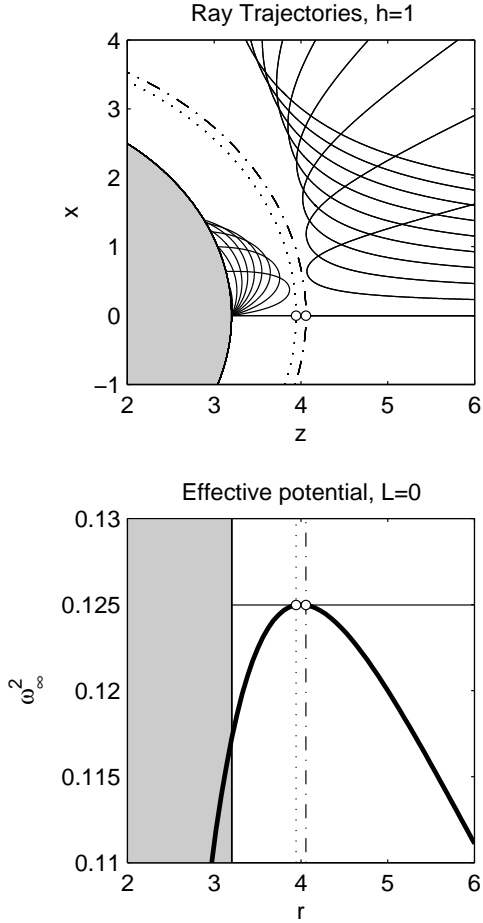}
\caption{Examples of ray tracing in the APW. Top: rays emitted from the surface 
of a CO (grey disc) and passing nearby with frequency 
$\omega_\infty=0.9999\omega_{\infty 0}$. We assume a plasma distribution with 
$h=1$. The absorption points for radial rays are marked with open circles, and 
the inner 
turning radius at $r_{-}$ is marked by a dotted circular arc. A variety of 
other rays are shown with angular momentum that increases in increments of 
$0.05$, to 
a maximum of $L=0.45$. These rays are emitted from the CO surface and reach a 
maximum distance $r \leq r_{-}$, before being reflected back to the surface of 
the CO. The exterior rays that pass by the CO have impact parameters defined as 
$b=L/ \omega_\infty$, and $r \geq r_{+}$. These rays are reflected away from 
the CO and return to infinity. The outer absorption radius $r_{+}$ (equation 
\ref{rpm}) is reached by the fiducial ray with $b=0$ and is marked as the 
dash--dotted circular arc. The absorption points for the fiducial rays are 
marked as open circles. Bottom: the effective potential (asymptotic plasma 
frequency) 
for the fiducial rays ($L=0$) is shown as the thick black line, and the inner 
and outer absorption radii marked as vertical dotted and dash--dotted lines, 
respectively. The points where absorption occurs are open circles and the CO 
interior is shaded grey. This example makes use of a highly relativistic star 
with $R/r_{\text{g}}=1.60$, $R=3.2M$ and a plasma frequency of the form 
$\omega_{\text{e}}^2 \propto 1/r$ ($h=1$).}
\label{fig:traj_pot}
\end{figure}

We demonstrate the behaviour of trapped and deflected rays in the APW in Fig. \ref{fig:traj_pot}. In this and the remainder of the examples in this work we 
assume $h=1$ and use $R/r_\text{g}=1.6$ with $R=3.2M$. We assume a ray 
frequency near the ceiling of the APW, $\omega_\infty = 0.9999 
\omega_{\infty0}$. 
The top panel of this figure illustrates trajectories that are emitted from the 
CO surface with $L$ varying from $0$ to $0.45$ in increments of $0.05$. The 
fiducial radially directed rays with $L=0$ reach $r_-$, marked as a dotted 
circle and are absorbed. The exact position of the absorption point of the 
fiducial ray is marked as an open circle. Rays with finite $L$ reach a 
maximum altitude and turn back to the CO. The exterior rays which are turned 
away from the CO are shown with impact parameters $b = L / \omega_\infty$, 
using the corresponding 
angular momenta for the interior rays. These trajectories turn away from the 
CO. The closest approach for the fiducial ray $r_+$ is marked as a dash--dotted 
circle, and the absorption point is explicitly marked for the external fiducial 
ray. The other rays are turned away from the CO and return to infinity.

The effective potential for the fiducial rays is shown in the lower panel of 
Fig. \ref{fig:traj_pot} as the thick curved line. The fiducial rays are 
represented by the horizontal thin lines, with the absorption points $r_\pm$ 
marked as open circles, and the corresponding radii marked with vertical dotted 
and dash--dotted lines. The interior of the CO is shown in both 
panels by the light grey area. Thus, when observed at 
frequencies within the APW ($\omega_\infty \leq \omega_{\infty0}$), the CO is 
effectively `cloaked' from detection by distant observers.

\subsection{Maximum impact parameter for rays within the EW}
\label{sub:impact}
The maximum impact parameter for rays at and above the high-frequency end of 
the EW is found by setting $r=R$ in equation \ref{bDef},
\begin{equation}
b_\text{max}=\frac{R n(R)}{A(R)^{1/2}}
\label{bDefMax}
\end{equation}
with emission angle $\delta_\text{max}=\upi/2$ for a ray that makes a grazing escape 
from the CO surface. However, for COs sheathed in power-law distributions of 
plasma with $h<2$, frequencies in the EW have the maximum impact parameter 
reduced since $r_\text{c}>R$. At $\omega_{0}$, the 
maximum impact parameter for rays to escape vanishes entirely, and the image of 
the stellar surface to a distant observer appears to be a single point. We call 
the maximum impact parameter in the EW $b_\omega$.

In general, the maximum impact parameter is the limiting case for which 
equation \ref{turningPtDef} has a single solution. A ray with maximum impact 
parameter ($\delta=\upi/2$) will enter the circular orbit at $r_\text{c}$. Thus, 
the maximum impact parameter is found by evaluating equation \ref{bDef} at 
$r=r_\text{c}$,
\begin{equation}
b_\omega=\frac{r_\text{c}n(r_\text{c})}{A(r_\text{c})^{1/2}}.
\label{bOmega}
\end{equation}
However, this expression presents a difficulty since solving for $r_\text{c}$ 
from the potential derivative requires the angular momentum $L$, which itself 
is limited by $b_\omega$ for escaping rays at a particular frequency. Therefore, 
we propose to use 
the 
turning point relationship (equation \ref{turningPtDef}) directly to discover 
the maximum impact parameter given only $\omega_\infty<\omega_{\infty+}$ and 
without a priori knowledge of $r_\text{c}$ explicitly.

For general $h<2$, the solution of the turning point equation 
must be found 
using numerical methods since terms with fractional exponents will be present. 
Analytical solutions exist for integer values of $h$, however the homogeneous 
$h=0$ case simply reproduces the behaviour of a massive particle with mass 
$m=\sqrt{k}$ \citep{kulsrudLoeb}, though the general procedure used below can 
also provide the analytical solution. For the remainder of this section we 
examine the $h=1$ case, for which equation \ref{turningPtDef} reduces to a 
cubic 
with the general form
\begin{equation}
r^3 +cr^2 +dr +f=0.
\label{cubicEq1}
\end{equation}
A unique solution for this equation can be found by enforcing the discriminant 
to vanish. This discriminant condition results in a second cubic equation in 
$b^2$, solved by standard methods \citep{cubicEqs}. To state the solution 
explicitly, we define 
\begin{equation}
C = \frac{k}{\omega_{\infty}^2}\left( 3M+\frac{k}{4 \omega_{\infty}^2}\right) 
-27M^2
\end{equation}
\begin{equation}
D = \frac{M k^2}{\omega_{\infty}^4} \left( \frac{k}{\omega_\infty^2} - 6M 
\right)
\end{equation}
\begin{equation}
F=\frac{M^2 k^2}{\omega_\infty^6}\left( \frac{1}{\omega^2} - 8M k \right)
\end{equation}
along with 
\begin{equation}
p=D-\frac{C^2}{3}
\end{equation}
and
\begin{equation}
q=\frac{2C^3 -9CD +27F}{27}.
\end{equation}
In terms of these quantities, the maximum impact parameter that causes the 
discriminant to vanish is
\begin{equation}
b_{\omega}=\sqrt{t_0-\frac{C}{3}}.
\label{cubicSol}
\end{equation}
When $p \neq 0$ the function $t_0$ is given in terms of hyperbolic functions 
\citep{cubicEqs}. For $4p^3+27q^2>0$ and $p<0$,
\begin{equation}
t_0=-2\frac{q}{|q|}\sqrt{-\frac{p}{3}}\cosh\left[\frac{1}{3}\cosh^{-1}
\left(\frac{-3|q|}{2p}\sqrt{-\frac{3}{p}}\right)\right]
\end{equation}
and for $p>0$,
\begin{equation}
t_0=-2\sqrt{\frac{p}{3}}\sinh\left[\frac{1}{3}\sinh^{-1}\left(\frac{3q}{2p}\sqrt
{\frac{3}{p}}\right)\right].
\end{equation}
With the maximum impact parameter we find the corresponding angular 
momentum for a given frequency, $L=\omega_\infty b_\omega$, and the critical 
radius $r_\text{c}$ follows. Numerical evaluation confirms that equations 
\ref{cubicSol} and \ref{bOmega} produce identical results.

\begin{figure*}
\includegraphics[bb=  126 315 484 484, clip, scale=0.88]{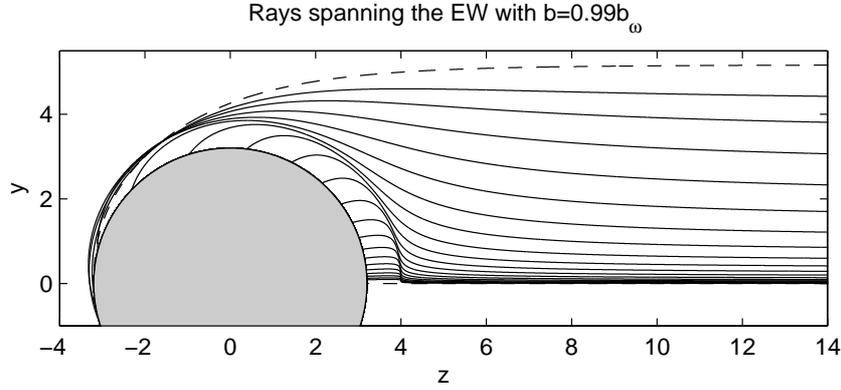}
\caption{Examples of ray tracing in the EW ($\omega_{0} < \omega \leq 
\omega_{+}$). Rays are emitted from the surface of a CO (grey disc). At a given 
frequency $\omega$, only rays with impact parameter less than the maximum 
$b<b_{\omega}$ escape the CO surface. At exactly $b=b_{\omega}$, rays enter the 
circular orbit at $r=r_\text{c}$. Therefore, we arbitrarily choose rays with impact 
parameter slightly below the maximum, $b=0.99 b_{\omega}$. The limiting cases 
are the rays marked by dashed lines at $b=0$ and the vacuum maximum, $b_\text{max}$. 
Low-frequency rays have a more apparent kink in their trajectories than the 
high-frequency rays at $r_\text{c}$. This example makes use of a highly 
relativistic star with $R/r_{\text{g}}=1.60$, $R=3.2M$ and a plasma frequency 
of the form $\omega_{\text{e}}^2 \propto 1/r$ ($h=1$).}
\label{fig:impact_par}
\end{figure*}

The angle that a ray with the maximum impact parameter makes with respect to 
the normal direction at the stellar surface is
\begin{equation}
\delta_\omega=\sin^{-1} \left( \frac{b_\omega}{b_\text{max}} \right).
\end{equation}
This shows that near $\omega_0$, rays that escape have small 
$\delta_\omega$ and are close to the radial direction $\hat{\mathbfit{r}}$. As 
the observed frequency $\omega_\infty$ increases, rays arriving at the observer 
will leave the CO at increasing angles, with the maximum
$\delta_\text{max}=\upi/2$ at the top of the EW. We show rays with impact 
parameter $b=0.99 b_\omega$ for $25$ frequencies ranging logarithmically across 
the EW using the radial limit of $\omega_0$ (equation \ref{w0h}) to $\omega_+$ 
for $h=1$ in Fig. \ref{fig:impact_par}. 

\begin{figure}
\includegraphics[bb= 139 132 458 680, clip, scale=0.75]{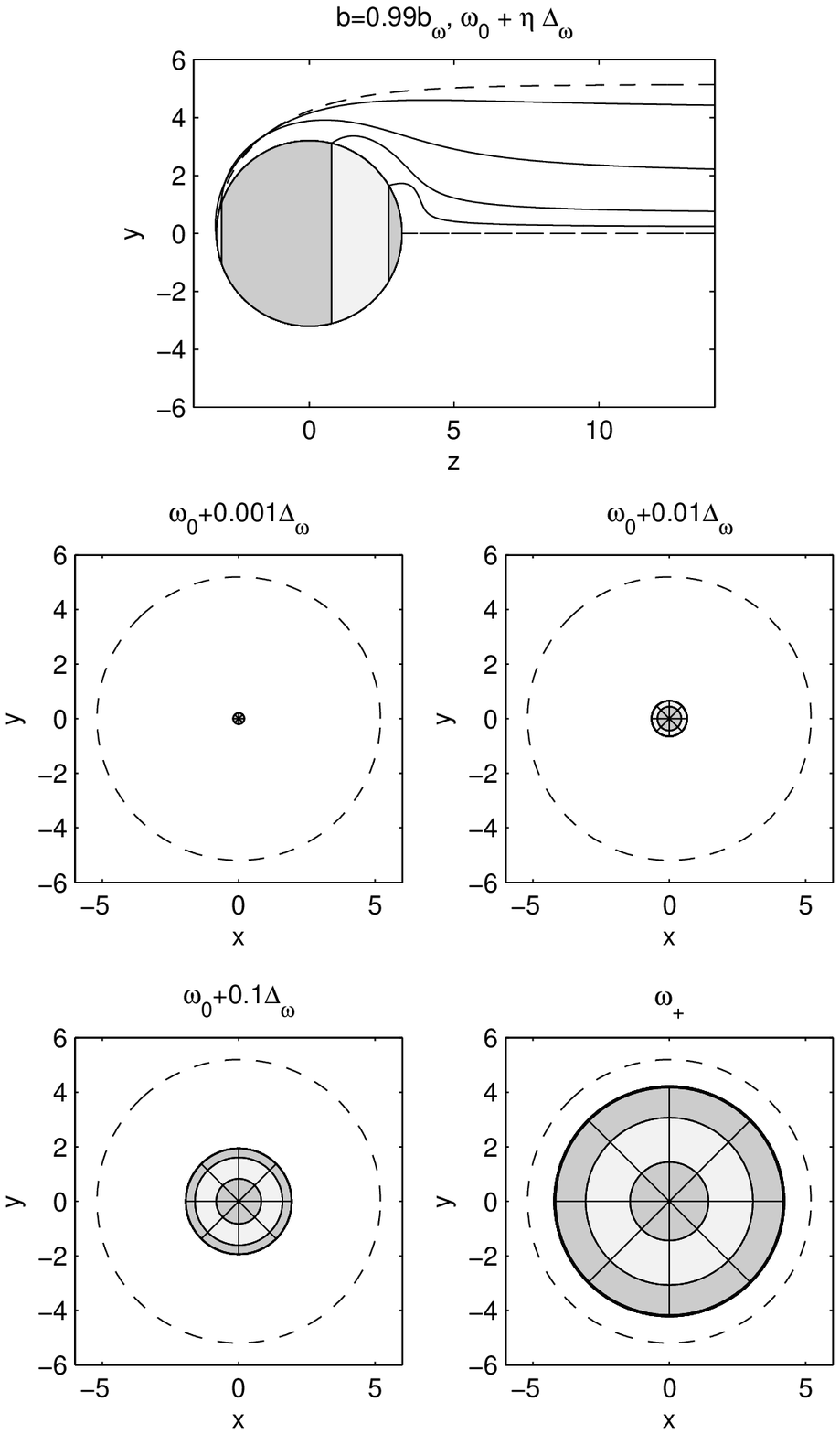}
\caption{Projection of CO surface elements to a distant observer. Top panel: 
rays with in the EW with frequency $\omega= \omega_0 + \eta \Delta_\omega$, 
where $\Delta_\omega=\omega_+ - \omega_0$ is the frequency range of the EW, 
with $\omega_0$ taken for the fiducial ray. We set $\eta$ to $10^{-3}$, 
$10^{-2}$, 
$10^{-1}$ and $1$. We then find the maximum impact parameter $b=0.99 b_\omega$ 
for each of these frequencies, and plot the path from the surface to the 
observer (in the $+z$ direction). The value of $\eta$ increases from the 
fiducial ray in the direction of increasing $y$. These impact parameters and 
the fiducial ray define the surface angle regions, which are represented as 
bands over the CO surface. Lower panels: each surface angle element is 
projected towards the observer and show the effect that observing frequency has 
on the appearance of the CO by changing the projections of the apparent solid 
angles. This example makes use of a highly relativistic star with 
$R/r_{\text{g}}=1.60$, $R=3.2M$ and a plasma frequency of the form 
$\omega_{\text{e}}^2 \propto 1/r$ ($h=1$).}
\label{fig:COsurf}
\end{figure}

We demonstrate how elements of the CO surface appear on the sky of a 
distant observer in Fig. \ref{fig:COsurf}. In the top panel of this figure, we 
launch rays from the surface of the CO to an observer located a great distance 
away in the $+z$ direction. For illustrative purposes we choose four 
frequencies in the EW, which we parametrize as $\omega= \omega_0 + \eta 
\Delta_\omega$, where $\Delta_\omega=\omega_+ - \omega_0$ is the width of the 
EW, and $\eta$ ranges from $0$ to $1$. For convenience, we will refer to these 
frequencies simply in terms of $\eta$. We choose $\eta$ to have the arbitrary 
values $10^{-3}$, $10^{-2}$, $10^{-1}$ and $1$, respectively. For each of these 
frequencies, we find the maximum impact parameter $b_\omega$ and consider the 
arbitrary impact parameter $b=0.99 b_\omega$ to avoid the circular orbit at 
$r_\text{c}$. These rays are illustrated in the top panel of Fig.
\ref{fig:COsurf} and increase in $\eta$ from the fiducial ray in the $+y$ 
direction. Let us then define the angle between the position of the ray on the 
stellar surface and the fiducial direction as the surface angle. To illustrate 
the effect that the plasma has on the appearance of a CO to a distant observer, 
we shade each of the surface angle elements in the top panel of Fig.
\ref{fig:COsurf} with an alternating light dark scheme. Due to spherical 
symmetry these surface regions define shaded rings on the surface of the CO 
concentric with respect to the line of sight. The surface angle regions contain 
all impact parameters $b \leq 0.99 b_\omega$. We plot these regions as a series 
of concentric discs on the lower 
panels. For each frequency, the surface angle regions are projected to the 
observer's sky. These plots show how the plasma distorts the view of the CO as a 
function of frequency.

For $\eta=10^{-3}$, the surface of the CO appears pointlike, and at 
$\eta=10^{-2}$ the large surface angle regions occupy a narrow range of large 
impact parameters and appear extremely compressed to the observer, subtending a 
small region at the limb of the CO disc. These areas subtend a larger solid 
angle on the sky as the frequency is increased. The $\eta=0.1$ and $\eta=1$ frequencies most clearly illustrate this point. Despite the maximum impact 
parameter for the $\eta=0.1$ and $\eta=1$ rays subtending nearly the same angle 
over the surface of the CO (i.e. in the top panel of Fig. \ref{fig:COsurf}), 
these regions project to different solid angles on the sky as seen in the 
bottom row of the figure. In these panels, the exterior dark grey ring 
represents the same region of the CO surface but it is markedly smaller in the 
$\eta=0.1$ 
panel with respect to the other solid angle elements. This shows that 
increasing the frequency does not simply scale the view of the CO surface to a 
larger 
size, but changes the proportions of the visible surface angle elements 
relative 
to 
one another. The demagnifying nature of the relativistic images produced by COs 
is well known and has been extensively studied in the lensing literature 
\citep{ve1, v1}, as well as in studies of the solid angles of NS surface 
elements projected to an observer in vacuum conditions \citep{pfc, dabrowski}.

\begin{figure}
\includegraphics[bb= 141 246 448 548, clip, scale=0.70]{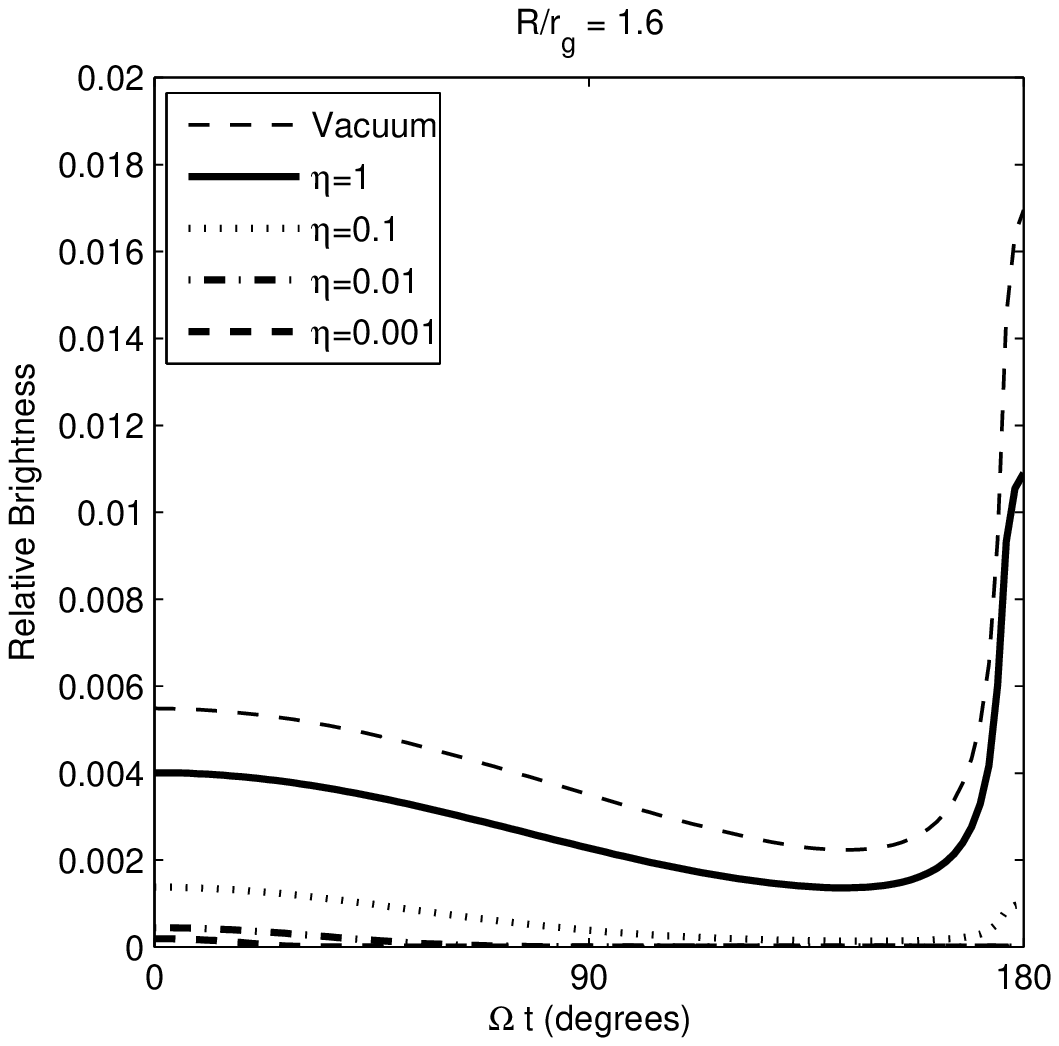}
\caption{Pulse profiles corresponding to the example shown in 
Fig. \ref{fig:COsurf}. We use a polar cap of angular size $5^\circ$ inclined at 
$90^\circ$ with respect to the rotation axis. The centre of the polar cap 
aligns with the line of sight at $\Omega t=0^\circ$ and is on the opposite side of the 
CO at $\Omega t=180^\circ$.}
\label{fig:lightCurve}
\end{figure}

We show the evolution of the pulse profile as a function of 
frequency in Fig. \ref{fig:lightCurve}. The pulse profile is calculated using 
the method from \citet{rogers2015}, with a single isotropically emitting 
radio-loud cap on the CO surface that has angular radius $5^\circ$ centred on 
the 
pole. The pole is set to an angle $90^\circ$ with respect to the rotation axis. 
This configuration aligns the line of sight and the centre of the polar cap at 
$\Omega t =0^\circ$, where $\Omega$ is the rotation frequency of the CO. The 
cap 
is on the opposite side of the star at $\Omega t=180^\circ$. All other 
properties are as in Fig. \ref{fig:COsurf}. The curves show the evolution of 
the pulse profile with frequency. The large peak that occurs when the polar cap 
is facing away from the viewer in the vacuum case is due to multiple imaging of 
the cap as rays are bent around the CO. In this configuration, the polar cap 
appears to the observer as a bright ring around the periphery of the NS. As the 
visible surface area decreases 
with frequency, the magnitude of this peak is correspondingly reduced. At the 
bottom of the EW, $\omega_0$, the pulse profile vanishes entirely. In addition 
to the 
frequency-dependent pulse morphology, the propagation of beamed pulsar 
radiation 
through plasma also results in a frequency-dependent time delay, which is 
strongly affected by the plasma properties \citep{rogers2015}.

\section{Discussion}
\label{sec:discussion}

In the analytical examples worked in the previous sections, we 
have assumed a CO with $R=3.2M$, slightly larger than the vacuum photon sphere at $r_\text{p}=3M$. 
Since we find the location of a turning point with $r_\text{c}>R$, the maximum 
possible compactness of a CO is relevant. To address the question of how 
compact a CO can be, we examined parametrized analytical fits to $22$ separate 
equations of state (EoS) for neutron matter \citep{gungorEksiEoS} that include a large cross-section of 
unique physics ranging from conventional npe$\mu$ matter to hybrid stars that 
possess exotic baryon condensates in their cores. For each of these EoS fits, we solve the 
Tolman--Oppenheimer--Volkov equations \citep{TOVEqns} to derive the mass--radius 
relationship. We also additionally include two examples for strange stars, 
based on perturbative QCD calculations \citep{strangeEoS, QSbagEoS}. The WFF1 equation of state \citep{WFF1-4, 
gungorEksiEoS} gives the smallest compactness ratio at $R/r_\text{g}=1.509$ above our lower limit of $1.5$ 
found from equation \ref{compressionRatio}. This implies that highly compact NSs that are sheathed 
in power-law distributions of plasma can allow for the frequency effects 
studied in Section \ref{sec:theory} provided that the plasma density drops off 
slowly enough ($h<2$). The soft EoS that produces COs with $R/r_\text{g}>2$ requires $h<1$ to show joint effects from gravitational 
lensing and plasma. While our results focus specifically on the power-law case, 
the frequency windows exist for any plasma distribution with density that 
drops off sufficiently slowly.

The $h=1$ case has particular relevance for the study of pulsar wind nebulae. 
An aligned rotating NS produces a particle wind with a density that is 
proportional to $1/r$ \citep{pwn1, pwn2, pwn3, pwn4}. Provided radio emission 
from low altitudes in the pulsar magnetosphere \citep{rad2, rad1}, both general 
relativistic effects and the refractive index of the magnetospheric plasma may 
be relevant. These effects may also be significant for models of eclipsing 
binary NSs or NS-black hole systems that are nearly edge-on with respect to the 
observer \citep{binary1, binary2}.

Finally, let us consider the ratios of the upper and lower EW limits with 
respect to the plasma frequency. For the cases with $h>2$, the asymptotic value 
of the plasma frequency is calculated with respect to the stellar surface. 
Evaluating equation \ref{asymPlasmaFreq} at the stellar surface gives 
\begin{equation}
\hat{\omega_{\infty \text{e}}}=\left[A(R)\frac{k}{R^h}\right]^{1/2}.
\end{equation}
However, when $h<2$, the circular orbit radius is external to the stellar 
surface, $R<r_\text{c}$. The propagation condition requires that observable 
rays 
have frequency in excess of the asymptotic plasma frequency $\omega_{\infty0}$ 
in order to escape the CO. Thus, for $h<2$ the asymptotic plasma frequency must 
be evaluated with respect to the circular orbit radius $r_\text{c}$ at the 
potential maximum. Since the most significant observable effects due to the 
plasma occur within the EW, it is of interest to compare the enhancement of the 
EW limits with respect to the asymptotic plasma frequency evaluated at the 
surface of the CO. Using equations \ref{w+} and \ref{asymPlasmaFreq} we find
\begin{equation}
\frac{\omega_{\infty+}}{\hat{\omega_{\infty 
\text{e}}}}=\left[\left(1-\frac{h}{2}\right)\frac{R-2M}{R-3M}\right]^{1/2}.
\label{upperRat}
\end{equation}
For a radially directed ray, we use equation \ref{w0h} to find
\begin{equation}
\frac{\omega_{\infty0}}{\hat{\omega_{\infty \text{e}}}}=\left[ \frac{1}{h+1} 
\frac{1}{A(R)} \left( \frac{h}{h+1} \frac{R}{2M} \right)^h \right]^{1/2}.
\label{lowerRat}
\end{equation}
Note that besides the mass and radius of the CO, the ratios depend only on the 
power-law exponent, such that the specific details of the plasma contained in 
the constant $k$ drop out entirely.

As an example, let us return to the case with $h=1$ and a compactness ratio 
$R/r_\text{g}=1.60$. With these parameters, we use equation \ref{lowerRat} to 
find the lower limit of the EW at $\omega_{\infty0}=1.033 \hat{\omega_{\infty 
\text{e}}}$ and equation \ref{upperRat} evaluates to 
$\omega_{\infty+}=1.732\hat{\omega_{\infty \text{e}}}$. In 
\citet{rogers2015}, we estimated an upper limit of the plasma frequency of the 
order of $\sim 100$ MHz, which is boosted to $\sim 170$ MHz from the effects 
presented in this work. Within this frequency range 
the pulse profiles should be drastically affected as shown in Fig. \ref{fig:lightCurve}. While the increase 
in the cutoff frequency for $h=1$ is modest, in the limit $h \rightarrow 0$ the 
limiting frequency ratio over the EW varies between 
$\omega_{\infty0}=1.633\hat{\omega_{\infty e}}$ and  $\omega_{\infty+}=2.450 
\hat{\omega_{\infty \text{e}}}$, a moderate boost compared to the plasma 
frequency cutoff at the stellar surface. Therefore, due to the increase 
in the frequency range of the EW from a potential maximum external to the CO 
surface, the combined action of plasma and gravitational effects may occur in 
COs at slightly higher frequencies than previously estimated in the 
literature \citep{review, rogers2015}. Despite this modification, the values above represent a rough estimate since relativistic 
plasma in a pulsar magnetosphere must have a plasma frequency 
$\omega_{e}<100$ MHz \citep{lowF1, lowF2, jones}.

The calculations in Section \ref{sec:theory} and the pulse 
profiles shown in Fig. \ref{fig:lightCurve} are a first approximation. In 
general, we expect other significant factors must also be taken into account to 
describe real COs. The assumption that pulsed radio emission occurs near the 
surface is most reasonable for millisecond pulsars with small magnetospheres 
\citep{gil, kijak} and rapid rotation that causes distortions of the pulse 
profile 
through the Doppler effect \citep{cadeau}. We have also treated the plasma as 
cold 
and ignored the magnetic field; however, both of these factors should play a 
significant role in describing the distribution of plasma within the 
magnetosphere and the propagation of electromagnetic radiation through it 
\citep{kulsrudLoeb, gedalin, petrova}. A complete description requires the full 
treatment of covariant radiative transfer, derived in \citet{brod03a} and 
\citet{brod04}. In fact, polarized emission can be used as a probe of the plasma 
density around X-ray binaries and pulsars \citep{brod03a}.

The narrow, low-frequency bands required to observe pulse 
profiles provide the greatest challenge to detecting
the effects described in this work directly. However, there are a number of 
instruments suited to studying the
low-frequency pulsed emission and time delays required, in particular the 
Low-Frequency Array \citep[LOFAR;][]{LOFAR}, which observes in the frequency range 
$10$--$240$ MHz \citep{obs3}, the Long Wavelength Array between $10$ and $88$ MHz 
\citep[LWA;][]{LWA}, the Ukrainian T-shaped Radio Telescope in the range of 
$10$--$30$ MHz \citep[UTR-2;][]{UTR2} and the Murchison Widefield Array 
between $80$ and $300$ MHz \citep[MWA;][]{murchison}. Extremely low frequency 
observations ($<30$ MHz) are complicated by a number of effects, including 
scattering from the interstellar medium, radio frequency interference from man-made sources and transmission through the Earth's 
atmosphere. Below $10$ MHz radio signals are completely attenuated by 
ionospheric propagation effects. Despite these difficulties, anomalous X-ray pulsars have been 
studied at $102$--$111$ MHz \citep{obs2, malovAXP}, and a large sample of 
pulsar light curves have been studied over a broad range of frequencies by LOFAR 
\citep{LOFAR_2, LOFAR_3}, down to as low as $15$ MHz \citep{LOFAR_1}. Generally, 
a low-frequency turnover in the spectrum of radio pulsars is associated with a 
corresponding evolution of the pulse profile, related to processes occurring near 
the rest-frame plasma frequency above the polar cap \citep{sieber, mm80}. While 
many NSs do not show evidence of such a turnover \citep{LOFAR_2, LOFAR_3}, this 
feature has been identified in a number of objects. For our purposes, we focus 
on the pulsars with a low-frequency turnover, such as PSR B1133+16 
\citep{PSRB1133}, the millisecond pulsars J2145--0750 \citep{PSRJ2145_1, PSRJ2145_2, PSRJ2145_3, 
PSRkuniyoshi15} and J1012+5307 \citep{PSR1012_1, LOFAR_3}, as well as a handful 
of additional systems \citep[see for example][]{PSRkuniyoshi15, LOFAR_3}. We do 
not suggest that lensing and plasma effects account for these observations in 
their entirety, but if components of the radio beam arise from emission near the 
surface of the NS, these components can be affected by joint effects of gravity 
and the surrounding plasma environment.

The Square Kilometre Array \citep[SKA;][]{SKA} is projected to 
be in operation by 2020. The sensitivity of the SKA will be vital for studying 
the emission properties of PSRs and detecting new systems. However, extreme 
measures are required to overcome the fundamental limitations of low-frequency 
observations imposed by artificial and ionospheric sources of radio 
interference. A proposed solution has been to place an antenna array composed of 
a swarm of satellites in a lunar orbit \citep{FIRST, DARIS, SURO, OLFAR}. In 
principle, such a space-based radio telescope could be sensitive to ultralow 
frequencies in the range $300$ kHz--$30$ MHz, opening a new avenue of inquiry 
into the emission processes active in NSs.

\section{Conclusions}
\label{sec:conclusions}
In the vacuum case, all frequencies of electromagnetic waves are affected 
equally by the lensing effects of the Schwarzschild metric. However, for a 
power-law distribution of plasma in the curved space--time around a CO, the 
issue of whether a star is observable at a particular frequency 
is more subtle, since the apparent radius of the star is frequency dependent 
with respect to a distant observer.

To quantify the behaviour of rays as a function of frequency, we defined two 
frequency windows for power-law plasma distributions with $h<2$: the EW is the 
window in which rays originating from the CO surface can escape to a distant 
observer, $\omega_0<\omega \leq \omega_+$, and the APW spans the frequency 
range $\omega_-<\omega \leq \omega_{0}$. In the APW, a family of rays exist 
that are emitted from and return to the CO surface. These behaviours are 
dependent on the presence of a plasma that drops off more slowly than the 
$r^{-2}$ contribution from the vacuum term in the effective potential. The 
frequency windows are found analytically for power-law distributions and have 
analytical expressions for integer $h<2$, but must be solved numerically for 
non-integer values. The maximum impact parameter for rays in the EW is given as 
a function of frequency from analysis of the classical turning points of the 
effective potential.

Finally, we studied plasma effects on the appearance of a CO formed on the sky 
of a distant observer. For a range of angles on the surface of a CO, we found 
the corresponding projection on the observer's sky. The solid angle subtended by the 
CO appears to change apparent size when the observing frequency is increased. 
At frequencies much larger than the upper limit of the EW, vacuum behaviour is 
recovered.

\section*{Acknowledgements}
It is my pleasure to acknowledge and thank Samar Safi-Harb for support through 
the Natural Sciences and Engineering Research Council of Canada (NSERC) Canada 
Research Chairs Program. I acknowledge Xinzhong Er for many interesting 
discussions and helpful comments that improved the clarity and flow of this 
manuscript, and thank Charlene Pawluck for proofreading and helpful 
suggestions for improving the text. I also acknowledge and 
thank the anonymous referee for providing valuable feedback that improved the 
manuscript.

\bsp	
\label{lastpage}
\end{document}